\documentclass[conference]{IEEEtran}
\usepackage{graphicx}
\usepackage{caption}
\ifCLASSOPTIONcompsoc
  \usepackage[caption=false,font=normalsize,labelfont=sf,textfont=sf]{subfig}
\else
  \usepackage[caption=false,font=footnotesize]{subfig}
\fi
\usepackage{indentfirst}

\usepackage{amsmath}
\allowdisplaybreaks[4]
\usepackage{amssymb}
\usepackage{times}
\usepackage{mathtools}
\usepackage{psfrag}
\usepackage{cite}
\usepackage{lastpage}
\usepackage{fancyhdr}
\usepackage{color}
 \usepackage{amsthm}
\usepackage{bigints}
\newtheorem{Lemma}{Lemma}

\newtheorem{lemma}[Lemma]{$\mathbf{Lemma}$}

\begin{document}
\title{New Designs of Robust Uplink NOMA in Cognitive Radio Inspired Communications\vspace{-0.6em}}
\author{\IEEEauthorblockN{Yanshi Sun\IEEEauthorrefmark{1},
Wei Cao\IEEEauthorrefmark{1}, Momiao Zhou\IEEEauthorrefmark{1} and Zhiguo Ding\IEEEauthorrefmark{2}}
\IEEEauthorblockA{\IEEEauthorrefmark{1}School of Computer Science and Information Engineering, Hefei University of Technology, Hefei, China}
\IEEEauthorblockA{\IEEEauthorrefmark{2} Department of Electrical Engineering and Computer Science, Khalifa University, Abu Dhabi, UAE, and\\ Department of Electrical and Electronic Engineering, University of Manchester, Manchester, UK.}
\IEEEauthorblockA{Email: \IEEEauthorrefmark{1}\{sys@hfut.edu.cn, caowei0115@163.com,mmzhou@hfut.edu.cn,\},
\IEEEauthorrefmark{2}zhiguo.ding@manchester.ac.uk\vspace{-1.6em}}}
\maketitle
\begin{abstract}
This paper considers a cognitive radio inspired uplink communication scenario, where one primary user is allocated with one dedicated resource block, while $M$ secondary users compete with each other to opportunistically access the primary user's channel. Two new designs of NOMA schemes, namely hybrid successive interference cancellation with power adaptation (HSIC-PA) and fixed successive interference cancellation with power adaptation (FSIC-PA), are proposed. The significant advantages of the proposed schemes are two folds. First, the proposed two schemes can ensure that the secondary users are opportunistically served without degrading the transmission reliability of the primary user. Besides,
the transmission robustness of the served secondary users can be guaranteed. Specifically, the outage probability error floors can be avoided for the secondary users, which is proved by asymptotic analysis in the paper. Extensive simulation results are also provided to demonstrate the superior performance of the proposed schemes.
\end{abstract}
	

\section{Introduction}
Non-orthogonal multiple access (NOMA) has obtained extensive attention in both academia and industry, due to its higher spectral efficiency and capability to support more users compared to conventional orthogonal multiple access (OMA) technique \cite{3GPPNOMAR13,3GPPNOMAR16,makki2020survey,fu2020zero,wang2019power}. The key idea of NOMA is to encourage multiple users to simultaneously occupy one channel resource block, which is not allowed in OMA. Consequently, how to address inter-user interference is one of key issues in NOMA aided
systems. To this end, a widely used method in NOMA to address inter-user interference is successive interference cancellation (SIC), where the users' signals are decoded in a successive manner \cite{higuchi2013non}.

The superiority of NOMA in future wireless communication network has been deeply investigated, as well as its
compatibility with other advanced technologies, such as multiple input multiple output (MIMO) \cite{choi2016power}, millimeter wave (mmwave) \cite{sysmmwave2018} and Terahertz (THz) communications \cite{zhang2020energy}, reconfigurable intelligent surfaces (RIS) \cite{wu2021coverage}, satellite communications \cite{lin2019joint} and so on. However, for uplink NOMA, there's an unfavorable feature that outage probability error floors exist in the existing schemes. The error floor means that the transmission reliability can not be arbitrarily high as transmit power increases, which significantly limit the application of NOMA in many practical scenarios. It was thought that the outage probability error floors are unavoidable in uplink NOMA. However, recent studies show that the conventional cognition is not correct. Specifically, a new design of SIC namely hybrid SIC (HSIC) is proposed for cognitive radio inspired uplink NOMA \cite{ding2021new,ding2020unveiling1,ding2020unveiling2}. In the proposed HSIC scheme, the decoding orders of users are dynamically determined according to the relationship between the instantaneous channel conditions and users' target rates. \cite{ding2021new,ding2020unveiling1,ding2020unveiling2} show that the proposed HSIC scheme can avoid outage probability error floors, under some constraints on users' target rates. The most important contributions of the series studies in \cite{ding2021new,ding2020unveiling1,ding2020unveiling2} are two folds.
First, \cite{ding2021new,ding2020unveiling1,ding2020unveiling2} show that it is possible to avoid outage error floors, at least under some specific conditions. Second, the work in \cite{ding2021new,ding2020unveiling1,ding2020unveiling2} indicates the importance of introducing HSIC to improve transmission robustness of uplink NOMA, while most existing work on NOMA applies fixed SIC order, either based on the channel state information (CSI) \cite{gao2017theoretical,zhang2016uplink} or quality of service (QoS) requirement of users\cite{zhou2018state,Dhakal2019noma}.

However, as mentioned above, the proposed scheme in \cite{ding2021new,ding2020unveiling1,ding2020unveiling2}  can only avoid outage probability error floors under some stringent conditions on users' target rates, which may not be met in many realistic scenarios. Thus, it is natural to ask the following two questions.
The first question is that is it possible to avoid outage probability error floors without any constraints on users' rates? And the second question is that is it necessary to apply HSIC to avoid outage probability error floors?

This paper aims to answer the above questions, by considering a cognitive radio inspired uplink NOMA scenario. In the considered scenario, one primary user is allocated with one dedicated channel resource block,
while there are $M$ secondary users who compete with each other to opportunistically share the primary user's resource block without degrading the outage performance of the primary user. Two new designs of NOMA schemes, namely HSIC with power adaptation (HSIC-PA) and fixed SIC with power adaptation (FSIC-PA) are proposed. Both schemes can avoid outage probability error floors without any conditions on users' target rates. The main contributions of this paper are listed as follows.
\begin{itemize}
  \item Two novel designs of uplink NOMA schemes are proposed, namely HSIC-PA and FSIC-PA\footnote{Note that the HSIC-PA scheme extends the scheme proposed in our previous work \cite{sun2021new} where only two users are considered, while the FSIC-PA scheme hasn't been proposed according to our best knowledge.}. In the proposed HSIC-PA scheme, the decoding order of the secondary user can be dynamically changed according to the channel conditions. While in the proposed FSIC-PA scheme, the decoding order of the secondary user  is fixed at the  second stage of SIC. Asymptotic analysis for the outage probabilities of the served secondary users achieved by the proposed two schemes are provided, which shows that both schemes can avoid outage probability error floors without any constraints on users' target rates. The fact that the proposed FSIC-PA scheme can avoid error floors indicates that HSIC is not necessary to avoid error floors.
  \item Numerical results are presented to demonstrate the superior performance of the proposed HSIC-PA scheme and FSIC-PA scheme, by comparing with the benchmark scheme termed HSIC-NPA proposed in \cite{ding2021new,ding2020unveiling1,ding2020unveiling2}. It is shown that FSIC-PA scheme performs better than HSIC-NPA scheme in the high SNR regime, but worse in the low SNR regime. Moreover, HSIC-PA scheme performs best among three schemes at all SNRs, which shows the importance of the combination of HSIC and PA in the design of uplink NOMA transmissions.
\end{itemize}
\section{System model}

  Consider an uplink NOMA communication scenario with one base station (BS), one primary user $U_0$ and $M$
secondary users $U_m$, $1\le m\le M$. Note that, in the considered scenario, ensuring the transmission reliability of $U_0$ is of the first priority, which has a preset target data rate denoted by $R_0$. In conventional OMA based schemes, the primary user is allocated with one dedicated resource block, which cannot be accessed by other users. While in the considered NOMA schemes of this paper, $M$ secondary users can compete with each other to opportunistically  access the channel resource block which is allocated to the primary user. Note that allowing secondary users to share the channel resource block of the primary user must be done in such a way that the QoS of the primary user $U_{0}$ is not degraded.

	The channel gain of the primary user $U_{0}$ is denoted by $g$, and the channel gains of the secondary users are denoted by $h_m$, $1\le m\le M$. In this paper, $g$ and $h_m$ are modeled as normalized Rayleigh fading gains, which means that $g$ and $h_m$ are independent and identically distributed (i.i.d) circular symmetric complex Gaussian (CSCG) random variables with zero mean and unit variance, i.e., $g\sim\mathcal{CN}(0,1)$ and $h_m \sim \mathcal{CN}(0,1)$. The transmit power of the primary user $U_{0}$ is denoted by $P_{0}$. The transmit power of the secondary user $U_m$ is denoted by $\beta P_s$, where
$\beta \in \left [0,1\right ]$ is the adjustable power adaptation coefficient of $U_{m}$, and $P_s$ is the maximum power of $U_m$. For simplicity, the background noise power is normalized to be $1$ throughout the paper.

In the rest of the paper, the $M$ secondary users are ordered according to their channel gains, i.e.,
\begin{align}\label{eq1}
		\left | h_{1} \right | ^2< \cdots <\left |  h_{M}\right|^2.
\end{align}

In this paper, two novel NOMA schemes are proposed, namely HSIC-PA scheme and FSIC-PA scheme.
It will be shown that both schemes can avoid outage probability error floors.
For each scheme, in each period of transmission, only the secondary user which can achieve the largest instantaneous achievable rate is allowed to transmit signal by sharing the primary user's resource block.
The proposed two schemes are described in the next two subsections.

\subsection{HSIC-PA Scheme}

To begin with, define an interference threshold denoted by $\tau (g)$ as follows:
	\begin{equation}
		\begin{aligned}
			\tau (g)=&\max\left \{ 0,\frac{P_{ 0}\left |g  \right | ^2 }{2^{R_{ 0}} -1} -1\right \}.
		\end{aligned}
	\end{equation}
Note that $\tau(g)$ can be interpreted as the maximum interference, with which $U_0$ can
still achieve the same outage performance as in OMA where the resource block is solely occupied by $U_0$. For more details on $\tau(g)$, please refer to \cite{sun2021new,ding2021new}.

For each secondary user $U_m$, its instantaneous achievable rate is determined by how its channel
gain compares to $\tau (g)$, which can be classified into the following two types:
\begin{itemize}
  \item Type I: the received signal power of $U_m$ at the BS is less than or equal to $\tau(g)$, i.e.,
  $P_{s}\left | h_{m} \right |^2 \le \tau (g) $. For this case, putting $U_m$ at the second stage of SIC
  can yield larger rate compared to putting $U_m$ at the first stage of SIC, and will not hinder the primary user from successfully decoding its signal. Thus, it is favorable to decode $U_m$'s signal at the second stage of SIC, and the achievable rate of $U_m$ is given by
	\begin{align}
		R_{\uppercase\expandafter{\romannumeral1}}^m=\log(1+P_{ s}\left | h_{ m}  \right |^2) .
	\end{align}
  \item Type II: the received signal power of $U_m$ at the BS is larger than $\tau (g)$, i.e., $P_{s}\left | h_{m} \right |^2 > \tau (g)$. For this case, the benchmark scheme termed HSIC-NPA which is proposed in \cite{ding2021new} only considers the case where $\beta$ is set to be $1$. Thus, to not degrade the QoS of $U_0$, $U_m$ can only be decoded at the first stage of SIC in HSIC-NPA, yielding the following achievable rate of $U_m$:
      \begin{align}
      R_{II,1}^m=\log(1+\frac{P_{s}\left | h_{m} \right |^2 }{P_{0}\left | g \right | ^2+1} ).
      \end{align}
  Note that the drawback of putting $U_m$ at the first stage of SIC is that, when $P_0|g|^2$ is large,  $R_{II,1}^m$ might still be small even with a large $P_{s}\left | h_{m} \right |^2$.
  To this end, the proposed HSIC-PA scheme offers an additional choice where $\beta$ can be set to be less than $1$ so that $\beta P_s|h_m|^2=\tau(g)$, which can provide opportunity to yield a larger achievable rate. As a result, $U_m$'s signal can be decoded at the second stage of SIC, yielding the following achievable rate of $U_m$:
  \begin{align}
		R_{\uppercase\expandafter{\romannumeral2},2}^m=\log(1+\tau(g)).
  \end{align}
Thus, in the proposed HSIC-PA scheme, when $P_{s}\left | h_{m} \right |^2 > \tau(g)$, the achievable rate of $U_m$ is given by:
  \begin{align}
  R_{\uppercase\expandafter{\romannumeral2}}^m=\max\left\{R_{\uppercase\expandafter{\romannumeral2},1}^m,R_{\uppercase\expandafter{\romannumeral2},2}^m\right\}.
\end{align}
\end{itemize}
According to the above discussions, the achievable rate of $U_m$ in HSIC-PA scheme can be concluded as:
\begin{align}
	R^m=\begin{cases}
		R_{\uppercase\expandafter{\romannumeral1}}^m,&P_{s}\left | h_{ m}  \right |^2 \le\tau(g) \\
		R_{\uppercase\expandafter{\romannumeral2}}^m,&P_{s}\left | h_{ m}  \right |^2 >\tau(g).
	\end{cases}
\end{align}
\subsection{FSIC-PA Scheme}
In this subsection, another scheme termed FSIC-PA is introduced.
Note that in HSIC-PA scheme, the secondary user's signal can be decoded either at the first or second stage of SIC. However, in FSIC-PA scheme, the secondary user can only be decoded at the second stage of SIC.

In FSIC-PA scheme, for each secondary user $U_m$, its instantaneous achievable rate can also be determined by considering the following two cases as in the last subsection.
\begin{itemize}
  \item Type I: the received signal power of $U_m$ at the BS is less than or equal to $\tau(g)$, i.e.,
  $P_{s}\left | h_{m} \right |^2 \le \tau (g) $. For this case,  it is as same as in the HSIC-NPA and the proposed HSIC-PA scheme, where $U_m$ is decoded at the second stage of SIC. Thus, the achievable data rate of $U_{m}$ is $\hat{R}^{m}_I =\log(1+P_{s}|h_{m}|^2)$, since the interference from $U_0$ can be removed by SIC.
  \item Type II: the received signal power of $U_m$ at the BS is larger than $\tau (g)$, i.e., $P_{s}\left | h_{m} \right |^2 > \tau (g)$. For this case, in the proposed FSIC-PA scheme, $U_m$ can only be decoded at the second stage of SIC. To carry out this strategy, $\beta$ is set to be less than $1$ so that $\beta P_s|h_m|^2=\tau(g)$. Thus, the achievable rate of $U_m$ for type II is $\hat{R}_{II}^m=\log(1+\tau(g))$
\end{itemize}

By concluding the above two cases, the achievable rate of $U_m$ in the FSIC-PA scheme can be expressed as:
\begin{align}
\hat{R}^m=\begin{cases}
	\hat{R}^m_{I},&P_{s}\left | h_{ m}  \right |^2 \le\tau(g) \\
	\hat{R}^m_{II},&P_{s}\left | h_{ m}  \right |^2 >\tau(g).
\end{cases}
\end{align}

Note that, the proposed HSIC-PA and FSIC-PA schemes can ensure that the outage performance of the primary user is the same as that in the OMA scheme, where the resource block is occupied by the primary user only.
Thus, this paper focuses on the performance of the opportunistically served secondary users.

\section{Asymptotic performance analysis for HSIC-PA and FSIC-PA}
Note that outage probability error floors were thought to be unavoidable in uplink NOMA. However, recent work proves that such cognition is wrong, by introducing the concept of hybrid SIC, which dynamically chooses the decoding order according to the users' channel conditions and quality of service requirements \cite{ding2021new,ding2020unveiling1,ding2020unveiling2}. Even so, the outage probability error floors can only be avoided under some stringent conditions on users' target rates in the proposed HSIC-NPA scheme in \cite{ding2021new,ding2020unveiling1,ding2020unveiling2}, which is not practical in many scenarios. In this section, it will be  proved  that both the proposed HSIC-PA and FSIC-PA schemes can avoid outage probability error floors without any constraints on users' target rates, which indicates the importance of power adaptation for improving transmission robustness of uplink NOMA.

Due to the limited space, this paper only focuses on the asymptotic performance analysis for HSIC-PA and FSIC-PA, which is helpful to understand the proposed two schemes.
The overall outage probability achieved by the served secondary users of HSIC-PA is defined as:
\begin{align}
 P_{out}=\text{Pr}\left(\max\{R^m, 1\leq m\leq M\}<R_s\right),
\end{align}
and that of FSIC-PA scheme is defined as:
\begin{align}
 \hat{P}_{out}=\text{Pr}\left(\max\{\hat{R}^m, 1\leq m\leq M\}<R_s\right),
\end{align}
where $R_s$ is the target rate of the secondary users.

For the ease of characterizing $P_{out}$ and $\hat{P}_{out}$, it is useful to define the event $E_m$, which denotes the event that there are $m$ users belonging to type I, particularly, $E_m$ can be expressed as follows:
		\begin{align}
			E_{m}\!=\!\begin{cases}
				\left \{ \left |h_{m} \right |^2< \frac{\tau (g)}{P_{s}},\left | h_{m+1} \right | ^2>\frac{\tau (g)}{P_{s}}  \right \},
				&1\le m\le M-1,\\
				\left\{|h_{1}|^2 > \frac{\tau (g)}{P_{s}}\right\}, & m=0, \\
				\left\{|h_{M}|^2 < \frac{\tau (g)}{P_{s}}\right\}, & m=M,
			\end{cases}
		\end{align}
where the extreme cases $E_0$ and $E_{M}$ denote the events where there is no type I secondary users and all the secondary users belong to type I, respectively.

The following two lemmas show that both HSIC-PA and FSIC-PA schemes can avoid outage  probabilities error floors.

\begin{lemma}
In the high SNR regime, i.e., $P_0\rightarrow \infty$, $P_s\rightarrow \infty$, the overall outage probability of the served secondary users in HSIC-PA scheme approaches zero.
\end{lemma}
\begin{IEEEproof}
	The outage probability of HSIC-PA $P_{out}$ can be rewritten as:
\begin{align}
				P_{out}=&\underbrace{\sum\limits_{m=1}^{M-1}P\left( E_{m},\max\left\{R^m_{\uppercase\expandafter{\romannumeral1}},R^M_{\uppercase\expandafter{\romannumeral2}} \right\}  <R_{s} \right)}_{Q_{1}}\notag \\
				&+\underbrace{P \left( E_{0},R^M_{\uppercase\expandafter{\romannumeral2}}<R_{s}\right)}_{Q_{2}}\notag\\
				&+\underbrace{P \left( E_{M},R^M_{\uppercase\expandafter{\romannumeral1}}<R_{s}\right)}_{Q_{3}}
\end{align}
In the following, it is shown that $Q_1$, $Q_2$ and $Q_3$ approach zero in the high SNR regime, respectively.

For $Q_{1}$, it  can be evaluated as follows:
	\begin{align}
		Q_{1}=&\sum\limits_{m=1}^{M-1}P\left( E_{m},\max\left\{R^m_{\uppercase\expandafter{\romannumeral1}},R^M_{\uppercase\expandafter{\romannumeral2}} \right\}  <R_{s} \right)\notag \\
		\overset{(a)}\le& \sum\limits_{m=1}^{M-1}P\left( R^m_{\uppercase\expandafter{\romannumeral1}}<R_{s} \right)\notag\\
	    =&\sum\limits_{m=1}^{M-1}P\left(\log(1+P_{s}|h_{m}|^2)<R_{s} \right) \notag\\
	    \overset{(b)}\le&(M-1)P\left(\log(1+P_{s}|h_{1}|^2)<R_{s}\right),
	\end{align}
where step (a) is obtained by following the fact that $R^m_{I}<R_{s}$ and $R^M_{II}<R_{s}$, and step (b) is obtained by noting that the users are ordered as in (\ref{eq1}).
It can be directly seen that $Q_{1}$ approaches zero in the high SNR regime. Therefore, $Q_{1}\to 0$ at hign SNR.	

Then, for $Q_{2}$, it  can be rewritten as:
	\begin{align}
		Q_{2}=&P \left( E_{0},R^M_{\uppercase\expandafter{\romannumeral2}}<R_{s}\right)\notag\\
		=&P\left( |h_{1}|^2>\frac{\tau(g)}{P_{s}},\max\left\{ R_{\uppercase\expandafter{\romannumeral2},2}^M,R_{\uppercase\expandafter{\romannumeral2},1}^M \right\}< R_{s}\right)  \notag\\
   	    \overset{(c)}=&\underbrace{P\left(|g|^2<\alpha_{0},\log(1+\frac{P_{s}|h_{M}|^2}{P_{0}|g|^2+1})<R_{s} \right)}_{Q_{2,1}}\notag\\
		&+\underbrace{P\left( |g|^2>\alpha_{0},|h_{1}|^2>\frac{|g|^2\alpha_{0}^{-1}-1}{P_{s}},\right.}_{Q_{2,2}}\notag\\
		&\underbrace{\left.\max\left\{ \log(|g|^2\alpha_{0}^{-1}),\log(1+\frac{P_{s}|h_{M}|^2}{P_{0}|g|^2+1})\right\} \!< \! R_{s}\! \right)}_{Q_{2,2}},
	\end{align}	
where $\alpha_{0}=\frac{2^{R_{0}}-1}{P_{0}}$
and step (c) is obtained by dividing the event into two cases, one
is $\tau(g) = 0$ and the other is $\tau(g) > 0$.
 For $Q_{2,1}$, it can be obtained that:
\begin{equation}
	\begin{aligned}
		Q_{2,1}
		\le	& P\left( |g|^2<\alpha_{0}\right) =1-e^{-\alpha_{0}}\!\to \! 0.
	\end{aligned}
\end{equation}
For $Q_{2,2}$,  it can be calculated as follows:
\begin{small}
	\begin{align}
	Q_{2,2}=	&P\left( |g|^2>\alpha_{0},|h_{1}|^2>\frac{|g|^2\alpha_{0}^{-1}\!-\!1}{P_{s}},\log(|g|^2\alpha_{0}^{-1})<R_{s},\right.\notag\\
		&\left. \log(1\!+\!\frac{P_{s}|h_{M}|^2}{P_{0}|g|^2+1}) <R_{s}\!\!\right)\notag\\
		&{\le} P\left(\log(|g|^2\alpha_{0}^{-1})<R_{s}\right)
		=1-e^{-2^{R_s}\alpha_{0}}\to 0,
	\end{align}
\end{small}
Therefore, it can be concluded that $Q_{2}\to 0$ at high SNR.

Finally, $Q_{3}$ can be evaluated as follows:
\begin{align}
	Q_{3}=&P\left(P_s|h_{M}|^2\!<\!\tau(g) ,\log\left( 1\!+\! P_s|h_{M}|^2\right) \!<\!R_{s},|g|^2\!>\!\alpha_{0}\right) \notag\\
	&=P\left(|h_{M}|^2<\frac{\tau(g)}{P_s},|h_{M}|^2<\alpha_{s}, |g|^2>\alpha_{0}\right)\notag\\
	&\le P\left( |h_{M}|^2<\alpha_{s}\right) =1-e^{-\alpha_{s}}\to 0,
\end{align}
where $\alpha_{s}=\frac{2^{R_{s}}-1}{P_s}$. Thus, $Q_{3}\to 0$ at high SNR .

Since $Q_{1}\to 0$, $Q_{2}\to 0$ and $Q_{3}\to 0$, we have $P_{out}\to 0$ in the high SNR regime, and the proof is complete.
\end{IEEEproof}

Recall that the benchmark HSIC-NPA scheme \cite{ding2021new,ding2020unveiling1,ding2020unveiling2} can avoid the outage probability error floor only if the following conditions apply: $\epsilon _{0}\epsilon _{s}<1$, where $\epsilon _{0}=2^{R_{0}}-1$ and $\epsilon_{s}=2^{R_{s}}-1$. However, the proposed HSIC-PA scheme can avoid error floor without any restrictions on the target rates of the primary user and the secondary users.
Thus, we can conclude that it is possible to avoid outage probability error floors without any constraints on users' rates.

\begin{lemma}
In the high SNR regime, i.e., $P_0\rightarrow \infty$, $P_s\rightarrow \infty$, the overall outage probability of the served secondary users in FSIC-PA scheme approaches zero.
\end{lemma}
	\begin{IEEEproof}
	The outage probability of FSIC-PA $\hat{P}_{out}$ can be rewritten as:
	\begin{align}
		\hat{P}_{out}=	&\underbrace{P(|h_{M}|^2<\frac{\tau(g)}{P_{s}},\log(1\!+\! P_{s}|h_{M}|^2)<R_{s},|g|^2\!>\!\alpha_{0})}_{F_{1}}\notag\\
		&+\underbrace{P(|h_{M}|^2>\frac{\tau(g)}{P_{s}},\log(1\!+\!\tau(g))<R_{s},|g|^2 \!>\!\alpha_{0})}_{F_{2}}\notag\\
		&+\underbrace{P(|g|^2<\alpha_{0})}_{F_{3}}.
	\end{align}
In the high SNR regime, it can be proved that $F_1$, $F_2$ and $F_3$ approach zero, as shown in the following.

First, by noting that $F_{1} \le P(\log(1+P_{s}|h_{M}|^2)<R_{s})$, it can be easily proved that $F_1$ approaches zero.
	
Then,  $F_{2}$ can be evaluated as:
	\begin{align}
		F_{2}=&P(|h_{M}|^2>\frac{\tau(g)}{P_{s}},\alpha_{0}<|g|^2<2^{R_s}\alpha_{0})	\notag\\
		\le& P(|g|^2<2^{R_s}\alpha_{0})=1-e^{-2^{R_s}\alpha_{0}}\to0.
	\end{align}

Finally,  $F_{3}$ can be evaluated as: 	
	\begin{align}
		&F_{3}=P(|g|^2<\alpha_{0})=1-e^{-\alpha_{0}}\to0.
	\end{align}

	Thus, the outage probability of FSIC-PA $\hat{P}_{out} \to 0$ in the high SNR regime. \end{IEEEproof}

The above asymptotic analysis shows that the proposed FSIC-PA scheme can also avoid outage probability error floors without any  constraints on $\epsilon _{0}$ and $\epsilon _{s}$, which indicates that it is not necessary to apply HSIC to avoid outage probability error floors. However, the HSIC-PA scheme performs better than FSIC-PA scheme as will be shown in the next section, which indicates the importance of the combination of HSIC and PA to improve the robustness of uplink NOMA.

\section{Numerical Results}
In this section, simulation results are provided to demonstrate the performance of the proposed HSIC-PA and FSIC-PA schemes. Comparisons with the benchmark HSIC-NPA scheme \cite{ding2021new,ding2020unveiling1,ding2020unveiling2} are also provided.
\begin{figure}[!t]
  \centering
  \vspace{0em}
\setlength{\abovecaptionskip}{0em}   % 调整图片标题与图片距离
\setlength{\belowcaptionskip}{0em}   % 调整图片标题与下文距离
\includegraphics[width=3.2in]{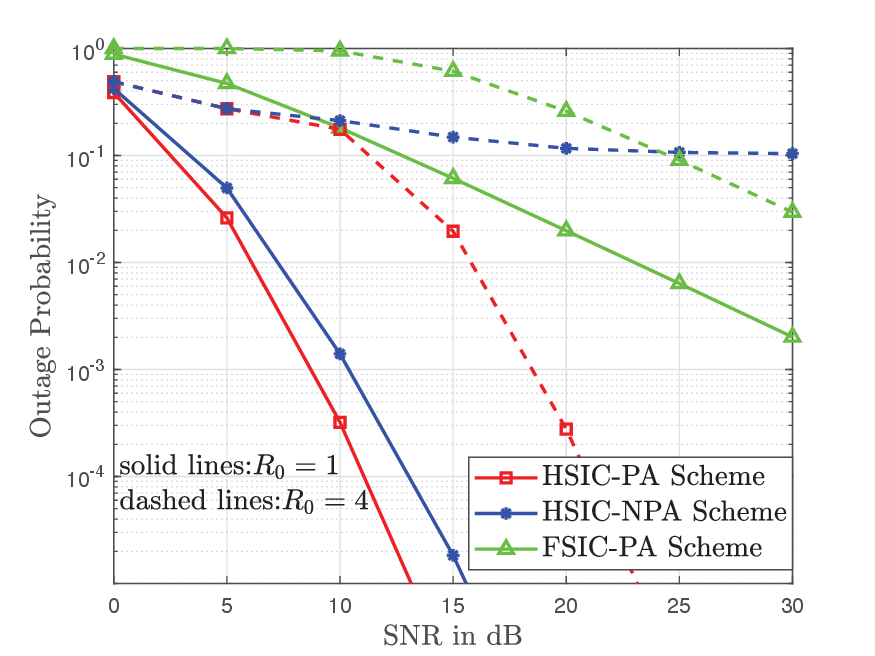}\\
\caption{Outage probabilities achieved by the HSIC-NPA, HSIC-PA and FSIC-PA schemes versus SNR, $M=4$, $R_{s}=1$ bit per channel use (BPCU), $P_{s}=P_{0}$.}
\label{outage_probability}
\end{figure}

Fig. \ref{outage_probability} shows the outage probabilities of the secondary users achieved by HSIC-NPA, HSIC-PA and FSIC-PA versus transmit SNR. As shown in the figure, for HSIC-NPA scheme, when $R_0=1$ BPCU, there is no outage probability error floor. However, when  $R_0=4$ BPCU, the outage probability error floor exists. This observation is consistent with the conclusions in \cite{ding2021new}, i.e., the error floor can only be avoided when $\epsilon_0\epsilon_s<1$. By contrast, the proposed HSIC-PA and FSIC-PA schemes can avoid outage probability error floors, since the outage probabilities achieved by both schemes continuously decrease as SNR increases. Fig. \ref{outage_probability} also shows that the HSIC-PA scheme performs best among the three schemes for all cases. However, FSIC-PA achieves larger outage probabilities than HSIC-NPA when $R_0=1$ BPCU, while for the case where $R_0=4$ BPCU, FSIC-PA performs better at high SNRs.
\begin{figure}[!t]
  \centering
  \vspace{0em}
\setlength{\abovecaptionskip}{0em}   % 调整图片标题与图片距离
\setlength{\belowcaptionskip}{0em}   % 调整图片标题与下文距离
\includegraphics[width=3.2in]{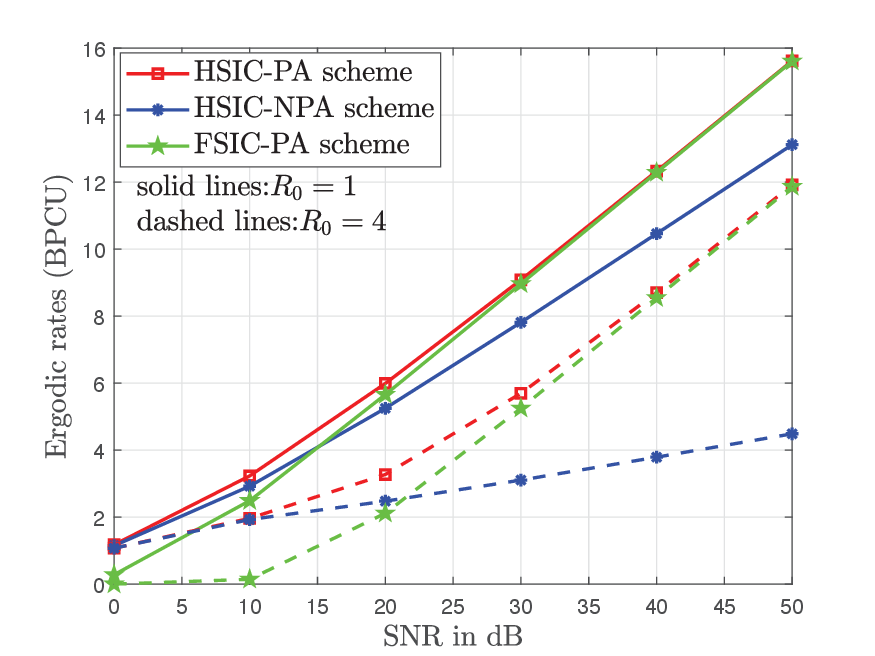}\\
\caption{Comparison among the three considered transmission schemes in terms of ergodic rates. $M=4$, $P_{s}=\frac{P_{0}}{3}$.}
\label{ergodic_rate}
\end{figure}

 Fig. \ref{ergodic_rate} shows the performance of the three schemes in terms of ergodic data rates achieved by the served secondary users.
 From the figure, it is shown that HSIC-PA scheme always achieves the largest ergodic rate among the three schemes, which is consistent with the observation in Fig. \ref{outage_probability}.
 Another interesting observation from  Fig. \ref{ergodic_rate} is that the performance of FSIC-PA approaches that of HSIC-PA in terms of ergodic rate at high SNRs, while the performance of HSIC-NPA approaches that of HSIC-PA in terms of ergodic rate at low SNRs. This observation indicates that it is preferable to set the
 secondary user at the first stage of SIC and use full transmit power at low SNRs, while it is preferable to set the secondary user at the secondary stage of SIC and use partial transmit power at high SNRs.

\begin{figure}[!t]
  \centering
  \vspace{0em}
\setlength{\abovecaptionskip}{0em}   % 调整图片标题与图片距离
\setlength{\belowcaptionskip}{0em}   % 调整图片标题与下文距离
\includegraphics[width=3.2in]{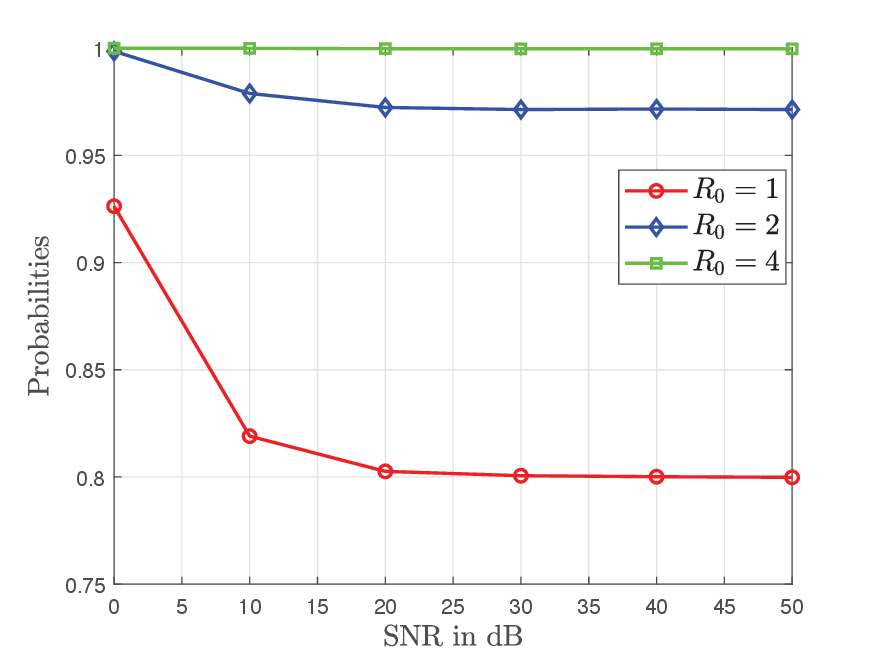}\\
\caption{The probabilities that the served secondary user belongs to type II. $P_{s}=P_{0}$, $ M=4 $.}
\label{Probability_type2}
\end{figure}

\begin{figure}[!t]
  \centering
  \vspace{0em}
\setlength{\abovecaptionskip}{0em}   % 调整图片标题与图片距离
\setlength{\belowcaptionskip}{0em}   % 调整图片标题与下文距离
\includegraphics[width=3.2in]{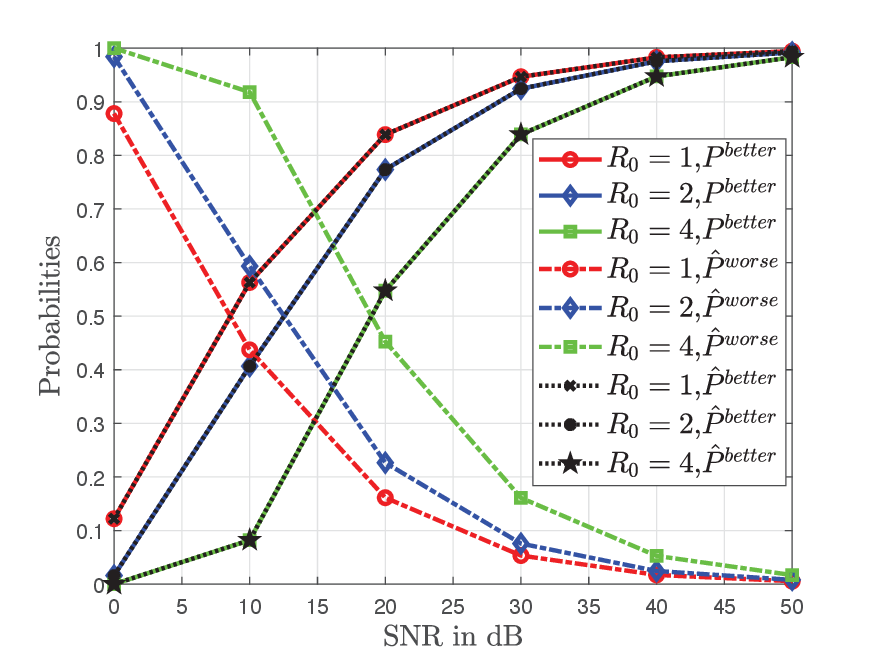}\\
\caption{$P^{better}$, $\hat{P}^{better}$ and $\hat{P}^{worse}$. $P_{s}=P_{0}$, $ M=4 $.}
\label{Probability_better}
\end{figure}

Fig. \ref{Probability_type2} and Fig. \ref{Probability_better} show a more detailed comparison of the proposed two schemes with the benchmark HSIC-NPA scheme. Note that, if the served secondary user belongs to type I, then the three schemes, i.e., HSIC-PA, HSIC-NPA and FSIC-PA, achieve the same instantaneous rate.
However, the three schemes differ from each other if the served secondary user belongs to type II.
Fig. \ref{Probability_type2} shows the probability that the served secondary user belongs to type II.
It is shown that as SNR increases, the probabilities converge to a constant.

When the served secondary user denoted by $U_{m^*}$ belongs to type II, then the achievable rate of $U_{m^*}$
can be denoted by ${R}_{II}$, $\hat{R}_{II}$ and $\bar{R}_{II}$ for HSIC-PA, FSIC-PA and HSIC-NPA scheme, respectively. For the comparison between HSIC-PA and HSIC-NPA, ${R}_{II}\geq\bar{R}_{II}$ always holds. Thus, it is sufficient to characterize the probability of the event that ${R}_{II}>\bar{R}_{II}$, which yields ${P}^{better}$ in Fig. \ref{Probability_better} defined by:
\begin{align}
{P}^{better}= \frac{P\left( \bar{R}_{\uppercase\expandafter{\romannumeral2}}<R_{\uppercase\expandafter{\romannumeral2}}, U_{m^*}\text{ is type II}\right) }{P\left(U_{m^*}\text{ is type II} \right) }.
\end{align}
By contrast, for the comparison between FSIC-PA and HSIC-NPA, $\hat{R}_{II}$ can be either larger or less than $\bar{R}_{II}$. Thus, it is necessary to consider both the probability that $\hat{R}_{II}>\bar{R}_{II}$ ($\hat{P}^{better}$ in Fig. \ref{Probability_better}) and
the probability that $\hat{R}_{II}<\bar{R}_{II}$ ($\hat{P}^{worse}$ in Fig. \ref{Probability_better}) simultaneously, which are defined as:
\begin{align}
\hat{P}^{better} =\frac{P\left( \bar{R}_{\uppercase\expandafter{\romannumeral2}}<\hat{R}_{\uppercase\expandafter{\romannumeral2}}, U_{m^*}\text{ is type II}\right) }{P\left(U_{m^*}\text{ is type II} \right) },
\end{align}
and
\begin{align}
	\hat{P}^{worse}=1-\hat{P}^{better},
\end{align}
respectively. An interesting observation from Fig. \ref{Probability_better} is that the curves for
$\hat{P}^{better}$ and ${P}^{better}$ coincide. Fig. \ref{Probability_better} also shows that $\hat{P}^{better}$ and ${P}^{better}$ increase with SNR, and approach $1$ in the high SNR regime. While
$\hat{P}^{worse}$ decreases with SNR and approaches $1$ in the low SNR regime.
The above observation can help to understand the phenomenon shown in Fig. \ref{outage_probability} and
 Fig. \ref{ergodic_rate}, and leads to the following suggestions for practical systems.
On the one hand, at high SNR, it is preferable to apply power adaptation and put the secondary user at the second stage of SIC. On the other hand, at low SNR, it is better to decode the secondary user at the first stage of SIC.

\begin{figure}[!t]
  \centering
  \vspace{0em}
\setlength{\abovecaptionskip}{0em}   % 调整图片标题与图片距离
\setlength{\belowcaptionskip}{0em}   % 调整图片标题与下文距离
\includegraphics[width=3.1in]{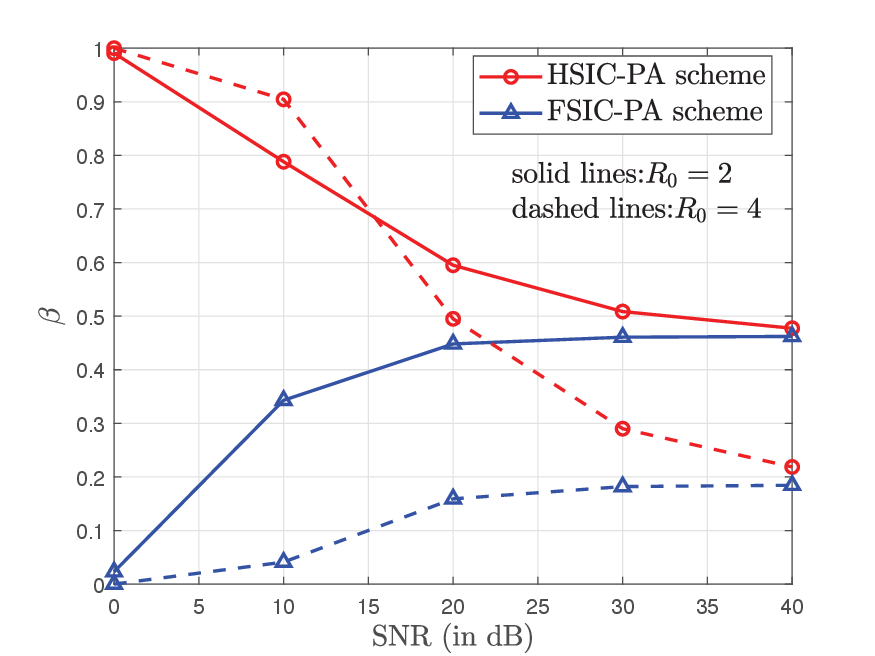}\\
\caption{Power consumption of the HSIC-PA and FSIC-PA schemes. $ M=4 $, $R_{s}=1$ BPCU, $P_0=P_s$.}
\label{power_consumption}
\end{figure}

Fig. \ref{power_consumption} shows the power consumption of HSIC-PA and FSIC-PA schemes. Note that  the HSIC-NPA scheme always chooses full power to transmit for the secondary users, i.e., $\beta$ is always set to be $1$, while $\beta$ can be set to be less than $1$ in the proposed HSIC-PA and FSIC-PA schemes. Thus, HSIC-NPA is more energy consuming than the proposed two schemes in this paper. From the figure, it can be observed that at low SNRs, $\beta$ approaches $1$ in HSIC-PA scheme and $\beta$ approaches zero in FSIC-PA scheme. Besides, as SNR increases, $\beta$ decreases in HSIC-PA scheme, while that in FSIC-PA scheme increases. More interestingly, the values of $\beta$ for both schemes approach a constant in the high SNR regime, which indicate that HSIC-PA scheme degrades to FSIC-PA scheme at high SNRs.

\section{Conclusion}
	In this paper, two novel NOMA schemes, namely HSIC-PA and FSIC-PA have been proposed, where  secondary users can opportunistically share the channel resource block with the primary user, without degrading the outage performance of the primary user compared to OMA.
The asymptotic analysis has been provided to characterize the outage performance of the proposed two schemes. It has been shown that both schemes can avoid outage probability error floors for the secondary user. Extensive numerical results have been provided to demonstrate the performance of the proposed schemes.
This paper has shown the importance of the decoding order of SIC and power control to improve the transmission reliability of NOMA.
\bibliographystyle{IEEEtran}
\bibliography{IEEEabrv,ref}
\end{document}